\documentclass{aastex}

\shorttitle{mJIVE--20: a VLBA survey for compact mJy radio objects}
\shortauthors{Deller, A.~T. \& Middelberg, E.}

\newcommand{\mjive}{mJIVE--20}

\begin{document}

\title{mJIVE--20: a survey for compact mJy radio objects with the Very Long Baseline Array}

\author{A.T. Deller\altaffilmark{1}, E. Middelberg\altaffilmark{2}}
\altaffiltext{1}{The Netherlands Institute for Radio Astronomy (ASTRON), Dwingeloo, The Netherlands}
\altaffiltext{2}{Astronomisches Institut der Ruhr-Universit\"at Bochum, Universit\"atsstra\ss{}e 150, 44801 Bochum, Germany}

\begin{abstract}
We present the description and early results of the mJy Imaging VLBA Exploration at 20 cm (\mjive).  \mjive\ is a large project on the Very Long Baseline Array which is systematically inspecting a large sample of mJy radio sources, pre--selected from the FIRST survey made with the Very Large Array, to identify any compact emission which may be present.  The survey is being undertaken using filler time on the VLBA, which utilizes short segments scheduled in bad weather and/or with a reduced number of antennas, during which no highly rated science projects can be scheduled.  The newly available multifield capability of the VLBA makes it possible for us to inspect of order 100 sources per hour of observing time with a 6.75$\sigma$ detection sensitivity of approximately 1 mJy/beam.  The results of the \mjive\ survey are made publicly available as soon as the data are calibrated.  After 18 months of observing, over 20,000 FIRST sources have been inspected, with 4,336 VLBI detections.  These initial results suggest that within the range 1 -- 200 mJy, fainter sources are somewhat more likely to be dominated by a very compact component  than brighter sources.  Over half of all arcsecond--scale mJy radio sources contain a compact component, although the fraction of sources which are dominated by milliarcsecond scale structure (where the majority of the arcsecond scale flux is recovered in the \mjive\ image) is smaller at around 30--35\%, increasing towards lower flux densities.  Significant differences are seen depending on the optical classification of the source.  Radio sources with a stellar/point--like counterpart in the Sloan Digital Sky Survey (SDSS) are more likely to be detected overall, but this detection likelihood appears to be independent of the arcsecond--scale radio flux density.  The trend towards higher radio compactness for fainter sources is confined to sources which are not detected in SDSS or which have counterparts classified as galaxies.  These results are consistent with a unification model of AGN in which less luminous sources have on average slower radio jets, with lower Doppler suppression of compact core emission over a wider range of viewing angles.

\end{abstract}


\section{Introduction}

The last decade has seen an accelerating trend towards astronomy
dominated by large surveys, in particular to study star formation and
accretion onto massive black holes on a cosmological scale. Sensitive
surveys, targeting large numbers of objects, are an indispensable
ingredient for this work. Radio surveys are an essential component of
these multi--wavelength studies due to their insensitivity to dust obscuration, and
because of their ability to detect non-thermal radiation from Active
Galactic Nuclei (AGN).

Identifying AGN is crucial for understanding galaxy evolution, since
their energetic feedback is increasingly understood to have a decisive
impact on star formation in their host galaxies, in particular when
they are radio emitters. Even if their activity cycles are 
short or intermittent, they can deposit sufficient energy in their
host galaxy's interstellar medium to suppress star formation \citep{croton06a, di-matteo05a}, but they can also
compress the interstellar gas via mechanical interactions and trigger
star formation \citep{klamer04a,gaibler12a}. However, identifying AGN is difficult, even with the most
comprehensive data sets. Nuclear activity can be shielded from our
view at any wavelength except towards the radio; spectroscopic methods
are observationally too expensive to be used on large scales with
hundreds or thousands of objects; and even radio surveys do not
normally provide sufficient information (spectroscopic or
morphological) to reliably identify AGN.

Unlike the aforementioned methods, radio observations using the Very
Long Baseline Interferometry (VLBI) technique have the ability to
make unambiguous AGN identifications. The high resolution requires brightness
temperatures of order $10^6$\,K for a detection to be made, and this
can only be reached in non-thermal sources. VLBI observations can
therefore play an important role in AGN identification. 

Bright VLBI sources with $S>100\,{\rm mJy}$ are so rare that
essentially all have been identified in VLBI calibrator searches
(e.g., the VCS campaigns; see \citealt{petrov08b} and references therein), and
significant numbers of faint sources at $S<1\,{\rm mJy}$ are beginning
to be detected by current wide-field VLBI observations of well-studied
extragalactic fields \citep{middelberg11a,middelberg13a}. 
However, the population of ``in--between" sources with flux densities of
1 mJy to 100 mJy have been comparatively ignored, primarily due to the observational difficulties. 
Surveys with hundreds of detections have lacked morphological and accurate flux density information \citep{porcas04a,
bourda10a}, whilst true imaging surveys have been restricted to much smaller samples \citep{garrington99a, garrett05a, wrobel05a, lenc08a}.
A large, comprehensive, and unbiased imaging survey of mJy sources is 
needed to bridge the gap between the wide/shallow and narrow/deep surveys, yielding input for
studies of galaxy evolution in the local ($z<1$) universe.

Our project, the mJy Imaging VLBA Exploration at 20 cm (\mjive),
aims to characterize the compact radio source population 
with flux densities between 1\,mJy and 100\,mJy by making high--resolution images of 
radio sources detected across 200+\,deg$^2$ in the Faint Images of the Radio Sky at
Twenty centimetres (FIRST) survey \citep{becker95a}. Its overlap with the {\it Sloan}
Digital Sky Survey \citep[SDSS;][]{york00a} ensures availability of photometry and in some cases
spectra, aiding interpretation of the data significantly.

By combining the data from this project with information on bright
sources from the VLBA calibrator surveys and information on faint
sources from deep field VLBI observations such as those of
\citet{middelberg13a}, we will also be able to construct the
differential source counts of radio-active radio sources over more
than 4 orders of magnitude in flux density, to study the evolution of
the AGN population with redshift and luminosity.

Another application of this dataset is the identification of 
radio sources which are suitable to be used as in-beam or 
nearby out--of--beam calibrators for other studies. 
The phase coherence of VLBI observations is typically
limited by the separation between calibrator and target, so it is
desirable to have a calibrator as close as possible to the target. 
 In particular, a dense grid of compact radio sources is increasingly
necessary for large astrometric campaigns, especially at lower 
frequencies \citep[e.g.][]{chatterjee09a,deller11b}. The \mjive\ catalogue
will greatly enhance the number of available calibrators
across the studied area, to the benefit of other VLBI observations made in
the surveyed region.

Finally, a large catalog of compact sources may prove useful as a starting point for searches for other exotic systems.  For example, multi--component systems could be inspected to search for binary AGN \citep[e.g.,][]{tingay11a,burke-spolaor11b}, which would be expected to present two compact cores with flat or inverted spectra.  Similarly, candidate gravitational lens systems could be identified from widely separated VLBI components with identical spectra which are offset from a bright elliptical galaxy.

We describe the \mjive\ survey in Section~\ref{sec:surveydesc} and the 
current source catalog in Section~\ref{sec:catalog}.  We detail some 
preliminary results extracted from the catalog data products in 
Section~\ref{sec:results}, and present our conclusions in 
Section~\ref{sec:conclusions}.

\section{Survey description}
\label{sec:surveydesc}

\subsection{Observations and scheduling}

The \mjive\ survey was approved with an initial allocation of 200 hours of observing time at the filler level priority (VLBA project code BD161), and has now been extended to a total of 600 hours (VLBA project code BD170).  In order to take advantage of VLBA filler time, it is necessary that the observations be relatively short, impose a limited burden on the recording media pool, and be tolerant to both bad weather and missing antennas.  The observing frequency was chosen to be 1.4 GHz to match the FIRST survey, and at this frequency weather conditions are unimportant.  Observations of 1 hour are sufficiently long to provide adequate $uv$\ coverage for source detections, although the ability to reliably reconstruct the structure of complicated sources is considerably reduced.  This disadvantage is particularly pronounced when several VLBA antennas are missing from the observation.  As the main purpose of this survey is the measurement of the peak flux density and approximate source size, however, this compromise is acceptable.

\begin{figure}
\begin{center}
\includegraphics[width=0.85\textwidth]{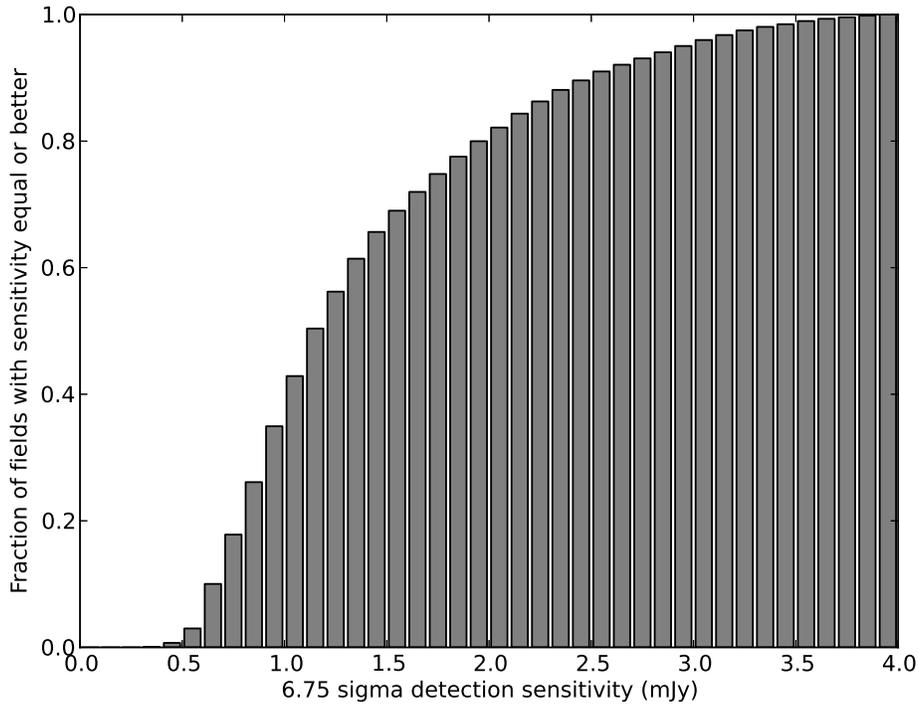}
\caption{A cumulative histogram of the detection sensitivity over all \mjive\ fields observed to date.  The median detection sensitivity is 1.2 mJy, but a small fraction of fields (which have been observed multiple times) can detect a compact source of peak flux density as low as 0.4 mJy.}
\label{fig:fieldsensitivities}
\end{center}
\end{figure}

\begin{figure}
\begin{center}
\includegraphics[width=0.85\textwidth]{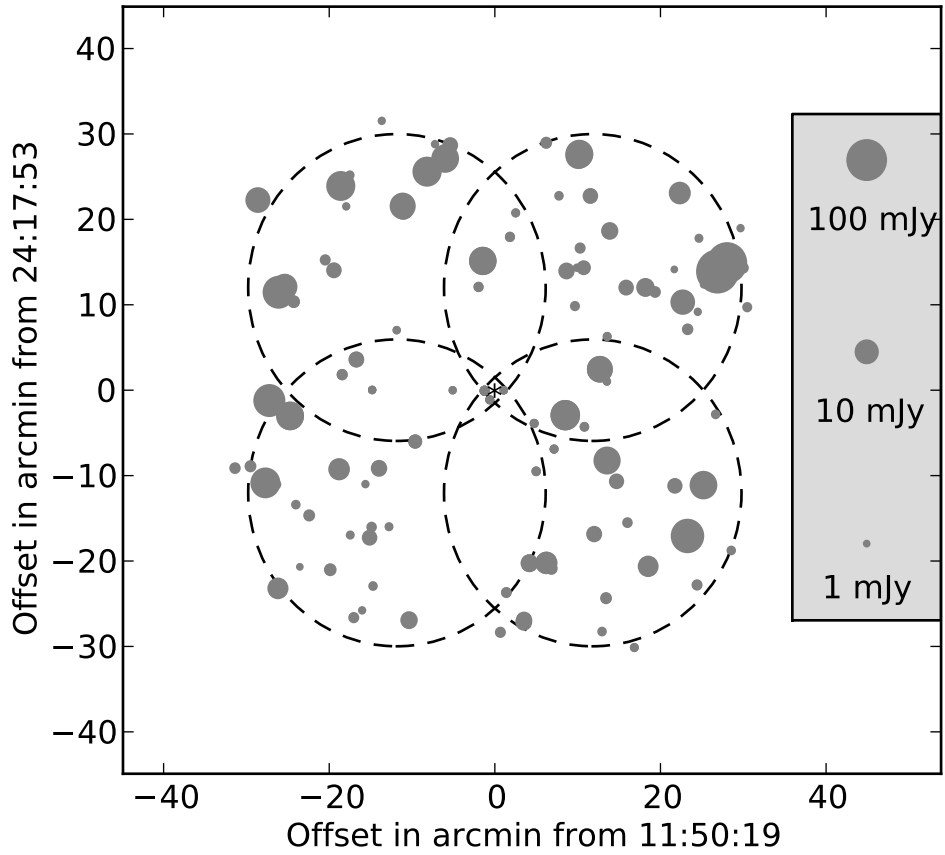}
\caption{An example of the pointing pattern layout for \mjive, showing the fields around J1150+2417.
The position of the calibrator source J1150+2417 is shown with an asterisk, and the 50\%
point of the primary beam response is shown with dashed lines.  FIRST sources which are targeted
with phase centers are shown as grey circles, with a radius proportional to the logarithm of their
FIRST peak flux density.}
\label{fig:pointings}
\end{center}
\end{figure}

A standard VLBA continuum observing setup (64 MHz of bandwidth in dual polarization, giving a total data rate of 512 Mbps) is used for these observations.  At this data rate, matching the point source sensitivity of the FIRST survey with the VLBA requires 15 minutes of on--source time.  Accordingly, 4 fields can be surveyed in each one hour observation.  To optimize the $uv$ coverage, pointings are observed in a round--robin fashion, spending 2 minutes on each pointing and repeating the entire loop 7 times.  In order to obtain a roughly uniform sensitivity over as wide an area as possible, the pointing centers are laid out with partial overlap, as shown in Figure~\ref{fig:pointings}.  In actuality, as the sensitivity at the edge of the field is considerably reduced, not every FIRST source would be detectable even if completely compact. Figure~\ref{fig:fieldsensitivities} shows a histogram of the detection sensitivity over all fields observed by \mjive\ to date -- the median detection sensitivity (with a detection threshold of 6.75$\sigma$, as described in detail in Section~\ref{sec:sourceid}) is 1.2 mJy.

Directly imaging the entire $\sim$200 square degrees observed by \mjive\ is barely feasible and certainly wasteful -- given the pixel size of 1 milliarcsecond demanded by the survey's angular resolution ($\sim$5 milliarcseconds), of order $\sim$10$^{15}$ pixels would be produced, along with hundreds of terabytes of intermediate data products.  Instead, the multi--field capability of the VLBA--DiFX software correlator \citep{deller07a,deller11a} is utilized to place phase centers at the location of all FIRST sources within 20\arcmin\ of the beam center (the 45\% point of the beam at the center bandwidth) for each pointing.  Since the FIRST positions are accurate at the arcsecond or sub--arcsecond level, these data products can then be heavily averaged, leading to a relatively low visibility data volume.

The multifield correlation uses an initial spectral resolution of 4 kHz, which is subsequently averaged (after the formation of the multiple phase centers) to 1 MHz.  The formation of the multiple phase centers takes place at a cadence of 100 Hz or faster, and the visibilities are then averaged to a time resolution of 3.2 seconds.  The reduction in image sensitivity (due to time smearing, bandwidth smearing and delay beam effects; see \citealt{morgan11a}) imposed by the multifield processing for a source at the edge of a given pointing is 20\%; considerable, but not significant compared to the primary beam amplitude attenuation of 55\%.  The amplitudes of the visibilities produced by DiFX are corrected for this decorrelation.  Some early \mjive\ observations used an 8 kHz initial spectral resolution and placed phase centers on sources within a 17\arcmin\ radius, and thus targeted fewer sources whilst suffering somewhat greater decorrelation for sources at the edge of the pointing.  After determining that the higher resolution did not place an undue burden on the correlator resources, this was expanded to the current setting.  The unprocessed visibility dataset size per source per pointing is 16 MB -- with an average of 35 FIRST sources per pointing, the total unprocessed visibility volume per 1 hour observation is approximately 2GB.  Using this multifield approach, all of the FIRST sources within an area of approximately 1 square degree (of order 100 sources) can be surveyed in one hour. 

Since the total extent of the FIRST survey exceeds 10,000 square degrees \citep{becker95a}, completely surveying the FIRST coverage with the VLBA is impractical at the current data rates.  Accordingly, when designing the survey we were free to select the subset of fields for which calibration will be simplest.  The most convenient calibration situation is when a calibrator is continuously available in the primary beam of all antennas.  This can be easily accomplished by centering the four fields shown in Figure~\ref{fig:pointings} on the chosen calibrator.  Calibrators were taken from the \verb+rfc_2012b+ catalog available at \verb+http://astrogeo.org/rfc/+, which makes use of a large number of historical VLBI calibrator surveys.  Since there are $\sim$1200 known VLBI calibrators in the area covered by FIRST, it is possible to cover a significant fraction of the FIRST area whilst continuously keeping a calibrator within the antenna primary beam.   This is the approach which has been taken by \mjive. Almost no calibrators have readily available 1.4 GHz information, as the calibrator search observations are typically carried out in dual frequency 2.3/8.4 GHz mode, and so suitable calibrators are selected on the basis of spectral index, brightness and apparent compactness from higher frequency observations.  Even when applying strict criteria to ensure that over 100 mJy of compact flux should be available at 1.4 GHz, over 200 calibrators remain.  At the current time, this supply of ``prime" calibrators has not yet been exhausted.  As \mjive\ observations continue, however, eventually all of these calibrators will be observed.  At that time, the strict calibrator criteria could be relaxed at the cost of small additional calibration overheads, and/or new bright sources identified in \mjive\ could be used as calibrators around which target fields can be placed.

Most \mjive\ fields have been observed only once to date.  This means that, depending on the number of antennas in the observation, a number of faint FIRST sources near the edge of the pointing grid (where the sensitivity is lower due to primary beam attenuation) may not be detectable even if completely compact.  When determining detection fractions (Section~\ref{sec:results}), the attained image rms and hence detectability of a source in each field is taken into account.  In general, more sources will be detected by observing a new field than by reobserving a prior field, which has driven our field selection to date.  However, re--observing fields allows a reliable estimation of the false positive ratio and missed source ratio at varying levels of detection significance, and so we have observed a number of fields a second time.  As the survey progresses, we will monitor the source statistics at varying flux levels and envisage eventually repeating many fields in order to probe the faintest FIRST sources (flux density $\sim$1 mJy) more deeply.

\subsection{The calibration and imaging pipeline}
\label{sec:pipeline}
The calibration and imaging pipeline for \mjive\ is a fully automated Parseltongue script \citep{kettenis06a}.  ParselTongue provides a python--based interface to classic AIPS\footnote{http://www.aips.nrao.edu/}, including calibration tables and the visibility data directly.  Many aspects of this script were taken from previous projects which searched large numbers of sources to identify compact inbeam calibrators for astrometric projects \citep[e.g.][]{deller11b}, but for completeness the entire pipeline is described below.

\begin{figure}
\begin{center}
\begin{tabular}{cc}
\includegraphics[width=0.45\textwidth]{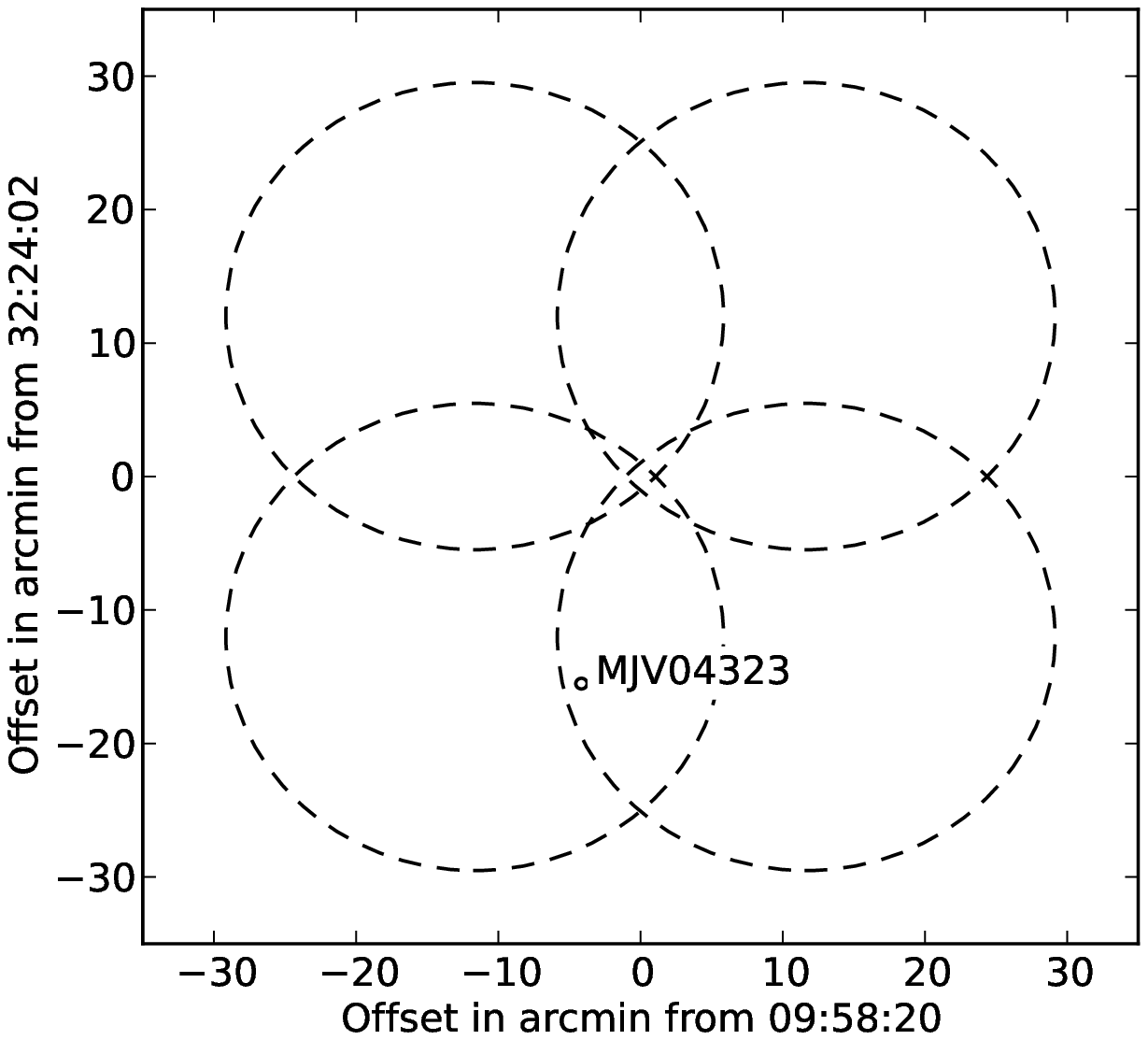} & 
\includegraphics*[width=0.375\textwidth]{MJV04323.clean.ps} \\
\includegraphics*[width=0.45\textwidth]{nopb_ovpt_amplitudes.ps} &
\includegraphics*[width=0.45\textwidth]{withpb_ovpt_amplitudes.ps}
\end{tabular}
\caption{\scriptsize{An example of primary beam correction for the source MJV04323.  The top left plot shows the position of MJV04323 in the pointing layout -- it is covered by two pointings, but is considerably nearer to the center of one pointing than the other.  The top right plot shows an image of the source made using all of the data after primary beam correction -- the peak flux is 37.2 mJy/beam and the noise is 0.20 mJy/beam.  Contours are drawn at 2, 4, 8, 16, 32 and 64\% of the peak flux density.  The bottom left panel shows the visibility amplitude on a subset of baselines before correction for primary beam effects.  Each grouping of visibility points is one scan, and the differing effects of primary beam attenuation on the alternating scans is clear, with the visibility amplitude being modulated by almost a factor of 2.  The bottom right panel shows the same set of visibility amplitudes after correction of primary beam effects - the improvement is obvious.  Imaging only the scans from the first pointing (where the source is closer to the pointing center) yields an image peak of 39.7 mJy/beam and a noise of 0.40 mJy/beam, while imaging only the scans from the second pointing yields a peak of 35.5 mJy/beam with rms 0.22 mJy/beam.  The difference in $uv$ sampling between the two pointings is likely partly responsible, but we conservatively estimate that the errors on the primary beam correction are on the order of 10\%.}}
\label{fig:beamcorrection}
\end{center}
\end{figure}

In multifield mode, the DiFX correlator produces independent FITS-IDI files containing the different phase centers.  The first FITS-IDI file contains the first phase center listed for each pointing center, the second FITS-IDI files contains the second listed phase center, and so on.  The following steps are undertaken in the pipeline:

\begin{enumerate}
\item The visibility datasets are loaded into AIPS and sorted into time order.
\item Additional flags are applied based on any user--provided information and on antenna elevation (data with an elevation less than 20\degr\ is flagged).
\item Ionospheric corrections are applied using the task TECOR.
\item Corrections are applied to update for the latest Earth Orientation Parameters (compared to those used at the correlator) using the task CLCOR.
\item Basic amplitude calibration is applied using the tasks ACCOR and APCAL; in each case, outliers in the derived amplitude corrections are clipped using the task SNSMO.
\item Amplitude corrections are applied for primary beam effects, using a ParselTongue script and a simple model of the VLBA beam (using a Bessel approximation to approximate the response of a uniformly illuminated 25 dish, and including the offsetting effects of the VLBA beam squint) to generate a calibration table directly.  A similar approach has been previously used by \citet{middelberg13a}.  Figure~\ref{fig:beamcorrection} shows an illustration of the effect of primary beam correction for one \mjive\ source.
\item Delay calibration is derived from the calibrator source using the AIPS task FRING and a 2 minute solution interval.  Polarizations and subbands are not averaged.  If available, a model of the source is used, otherwise a point source model is assumed.  Automated clipping of the solutions is applied with the task SNSMO.
\item Phase--only self--calibration is derived using a 30 second solution interval, coherently averaging all subbands and polarizations, using the task CALIB. A shorter solution interval can be used than in the preceding delay calibration step since more bandwidth is being combined, allowing short timescale atmospheric errors to be corrected.
\item Each source is split, applying all calibration.  The split datasets for sources which appear in multiple fields are concatenated using the task DBCON, ultimately leaving a single UV dataset per FIRST source.
\item The \mjive\ catalog is examined to determine if there are any bright VLBI sources near the current source which could potentially contribute confusing flux on some baselines.  If so, a ParselTongue script is run to flag baselines when the predicted decorrelation of the potential confusing source due to time and bandwidth smearing is insufficient to guarantee that there will be no artifacts generated in the image.  This confusion--flagging script is described in more detail in Section~\ref{sec:sourceid}.
\item For each target source, a wide--field image is generated using the task IMAGR with natural weighting.  The image size is 4096x4096 pixels, with a pixel size of 1 milliarcsecond, sufficient given the FIRST astrometric accuracy of 1\arcsec\ (at the flux limit, with 90\% confidence).  The peak pixel within the inner 90\% of the image is identified, and the peak value and image rms are recorded.  For sources which are ultimately found to be non--detections, this peak pixel value from the widefield image is used as the upper limit for the \mjive\ flux density.
\item The dataset is re--centered on the position of the peak pixel using the task UVFIX.  It is then averaged in frequency to a single spectral channel per subband (i.e., resolution 16 MHz) and in time to 20 seconds.
\item The predicted image rms is computed based on the sum of the visibility weights (described in more detail below).
\item The re--centered dataset is imaged using IMAGR, with a smaller size (1024x1024 pixels, pixel size 0.75 milliarcseconds).  As before, natural weighting is used.  A circular clean window is placed around the central pixel with a radius of 25 pixels, and the data is cleaned until a limiting rms of 0.2 mJy or 1000 clean components is reached.
\item Source detection and extraction is performed using the \verb+blobcat+ package \citep{hales12a}, with a detection threshold $T_d = 6.5$ and a flood threshold $T_f=5.0$ (see \citealp{hales12a} for an explanation of these quantities).  If blobcat detects one or more sources and the peak signal--to--noise ratio exceeds 6.75$\sigma$, then the peak flux, integrated flux over all components and the error in these quantities are all recorded for the \mjive\ catalog, otherwise the image peak residual and rms is recorded.
\item If the observed image rms is more than 1.33 times the predicted image rms, the pipeline is repeated from step 10 using a larger initial image (8192x8192 pixels).  This can catch cases where the noise is raised due to sidelobes from a VLBI source between 2 and 4 arcseconds from the FIRST position, which can occur rarely when the FIRST source is resolved and the compact component is not coincident with the peak of the low--resolution emission. 
\end{enumerate}

The calculation of expected image rms based on visibility weights is possible because weights in AIPS are nominally in units of 1/Jy$^{2}$.  The expected image rms in Jy when using natural weighting ($\sigma_{\mathrm{image}}$) is therefore simply the square root of the inverse of the summed weights $w_{ij}$:
\begin{equation}
\sigma_{\mathrm{image}} = \frac{1}{\sqrt{\sum{w_{ij}}}}
\end{equation}
 However, for VLBA data a correction factor is necessary, because the weights provided in the FITS file output of the VLBA correlator are initially completeness values (in the range 0.0 to 1.0) rather than true 1/Jy$^{2}$ values.  Accordingly, the weight sum needs to be corrected by the product of the original integration time (in seconds; equal to 3.2 for \mjive) and spectral resolution (in Hz; equal to 10$^6$ for \mjive).  AIPS corrects the weights for time and frequency averaging performed in after loading the data, which is why the necessary correction factor is defined by the original integration time and spectral resolution, not the final values.  The weights are extracted and summed using a ParselTongue script which traverses all of the visibilities in the dataset.  We find that for fields without a complex, confusing source (i.e., non--detections or point--like sources) the agreement between the predicted and observed image rms is very good; the mean value of the ratio is 1.02 and the standard deviation is 0.1.

At each stage where calibration is derived and applied, plots of the calibration solutions are generated with the task LWPLA.  These plots can be inspected after the pipeline completes and if necessary new flagging information can be provided and the pipeline rerun.  This step is necessary on less than 10\% of observations.  A complete pipeline run takes of order 30 minutes on a reasonably modern, low--end desktop machine (Intel Core2 Duo, 3.0 GHz, 2 GB RAM).

The averaging of the dataset to 16 MHz in frequency and 20 seconds in time reduces the data volume to around 300 kB per field per pointing, small enough to allow us to make all of the $uv$\ data available online at the \mjive\ site (\verb+http://safe.nrao.edu/vlba/mjivs/products.html+).  Time and bandwidth smearing mean that only a relatively small image can be made free of artifacts (around 1 square arcsecond), but since the dataset has already been centered on the peak pixel from the widefield map this is not an issue for the vast majority of sources.  Higher resolution datasets can be generated as needed to image sources with significant structure over larger separations.

\subsection{Source identification, parameter fitting, and completeness}
\label{sec:sourceid}

As noted above, the primary pipeline uses blobcat to model potential sources, and we impose a signal--to--noise threshold at 6.75$\sigma$.  If no pixels in the image exceed 6.75$\sigma$, it is considered a non--detection.  Since complex, resolved sources generally show a higher level of background noise, this threshold discriminates somewhat against complex sources, a bias which is partially addressed below.  At the time of publication, over 90 fields had been observed more than once, allowing us the opportunity to evaluate our detection threshold by identifying false positives (which appear in a single epoch but not in the combined data from multiple observations) and near misses (which are clearly detected in combined data and are seen at the same position in a single observation but with a signal--to--noise ratio just below the cutoff threshold).  Highly variable sources (see Section~\ref{sec:variable} below) might be responsible for a small fraction of ``false positives" but at such a low level as to not bias the results significantly.  We find that at a threshold of 6.75$\sigma$, a maximum of 0.3\% of all \mjive\ targets (1.6\% of all detections) are false positives.  Lowering the threshold to 6.5$\sigma$ would increase the number of \mjive\ detections by 1.25\%.  However, of every 5 new ``detections" in the range 6.5 -- 6.75$\sigma$, only 2 would be real sources, and the other 3 would be spurious.  Raising the threshold to 7$\sigma$ would reject at least 4 real \mjive\ sources for every spurious \mjive\ detection that is rejected, assuming the effects of source variability over short timescales to be negligible.  Accordingly, we set the detection threshold to 6.75$\sigma$.

Complex, resolved sources are difficult to identify in any automated pipeline.  This is particularly true for short VLBI observations, where the sampling of the $uv$\ plane is exceedingly sparse.  One simple method of doing so is to compare the noise in the image with expectations.  As noted above, a predicted image rms is produced for each source, based on the sum of the visibility weights.  Fields where the peak pixel considerably exceeds that expected from the expected image sensitivity can thereby be noted as probable complicated sources.  We have applied this procedure, noting fields where the peak pixel exceeds the expected image rms by a factor of 12 or more.  These sources (of which there are a relatively small number, less than 0.2\% of all targets) are then inspected by hand and rejected if spurious.  Non--spurious sources are entered into the catalog with the observed peak flux density values, and a special ``COMPLEX" flag in the catalog is set (see Section~\ref{sec:catalog} below).

Due to the high resolution and fringe rate typical of VLBI observations, sidelobe rejection is a much simpler problem than for observations with shorter--baseline instruments.  Time and bandwidth smearing will remove confusing sources in all areas except for their immediate vicinity, and the density of sources bright enough to be problematic is extremely low.  However, this time and bandwidth smearing makes it extremely difficult to accurately subtract these rare bright sources, and so as described in Section~\ref{sec:pipeline}, baselines are flagged when there is a {\em possibility} that a confusing source could contribute image artifacts in a target field.  The decision to flag is based on a worst case estimate which uses the envelope of the decorrelation function.  Such a worst--case approach is necessary because source subtraction is not feasible, since the exact decorrelation due to time and bandwidth smearing is difficult to calculate based on the information available in the FITS file. The confusing--source flagging pipeline selects all sources within 90\arcsec\ of the current target which have an integrated VLBI flux in excess of 50 mJy, as well as all sources within 300\arcsec\ with an integrated flux exceeding 1 Jy.  The time and bandwidth smearing are estimated based on the delay and rate as calculated from the baseline $w$ term.  Flags are written covering 10 minute timeranges if the maximum remaining correlated flux from the confusing source on the baseline exceeds 50 mJy (in most cases, however, the correlated flux will be much less, due to the worst--case assumptions used).  Approximately 2\% of all visibilities are flagged due to potential confusing sources, although the effects are dominated by a few equatorial fields with particularly bright calibrator sources.

\section{The \mjive\ catalog}
\label{sec:catalog}
The version of the \mjive\ catalog current at the time of publication, including data from 364 hours of VLBA observations made on or before MJD 56565 (2013 September 30) and covering 21,396 FIRST sources, is shown in Table~\ref{tab:catalog}. The \mjive\ catalog is updated regularly, and the latest machine-readable version is available for download at \verb+http://safe.nrao.edu/vlba/mjivs/catalog.html+.

Column 1 gives the \mjive\ source identifier name.  Columns 2 and 3 give the right ascension and declination obtained from the FIRST catalog.  Columns 4 and 5 give the FIRST peak and integrated flux density respectively in mJy/beam and mJy.  Columns 6, 7 and 8 give the VLBI synthesized beam major axis (mas), minor axis (mas) and position angle (\degr) respectively.  Column 9 gives the VLBI peak flux density (or limit) in mJy/beam.  Additional columns not shown in Table~\ref{tab:catalog} but available online are filled only for VLBI detections.  Column 10 gives the error on the VLBI peak flux density in mJy/beam, and columns 11, 12, 13 and 14 give the VLBI right ascension, error in right ascension, declination and error in declination respectively.  Column 15 gives the ratio of VLBI peak flux density to FIRST peak flux density.  Columns 16 and 17 give the integrated flux density in Jy from blobcat and its error.  Column 18 gives the ratio of VLBI integrated flux density to FIRST peak flux density.  Columns 19 and 20 give the deconvolved major axis of the single gaussian component fit and its error (max), columns 21 and 22 give the deconvolved minor axis and its error (mas), and columns 23 and 24 give the deconvolved position angle and error (degrees).  Column 25 contains the "COMPLEX" flag, denoting a source for which the signal--to--noise ratio of the single gaussian fit was less than the cutoff value of 6.75, but for which the peak image pixel exceeded a threshold set based on expected image noise, triggering a manual inspection of the data which verified this source to be real.

\begin{deluxetable}{cccccccccccccccccccccccccccccc}
\setlength{\tabcolsep}{0.02in} 
\tabletypesize{\tiny}
\tablecaption{Excerpt from the \mjive\ catalog}
\tablewidth{0pt}
\tablehead{
\colhead{\mjive} & \colhead{Right asc.} & \colhead{Decl.} & \colhead{Peak flux density} & \colhead{Int. flux density} & \colhead{\mjive\ beam} & \colhead{\mjive\ beam} & \colhead{\mjive\ beam} & \colhead{Peak flux density} \\
\colhead{Identifier} & \colhead{(FIRST)} & \colhead{(FIRST)} & \colhead{(mJy/beam, FIRST)} & \colhead{(mJy, FIRST)} & \colhead{major axis (mas)} & \colhead{minor axis (mas)} & \colhead{pos. angle (\degr)} & \colhead{(mJy/beam, \mjive)}
}
\startdata
MJV02915 & 00:14:33.699 & $-$00:13:21.04 & 7.8	 & 13.7	 & 16.5 & 6.2 & 6.4 & $<$1.32 \\
MJV02913 & 00:14:34.246 & $-$00:10:02.64 & 0.8	 & 0.8	 & 16.3 & 6.3 & 4.8 & $<$1.21 \\
MJV02910 & 00:14:34.284 & $-$00:07:10.12 & 2.5	 & 2.1	 & 16.3 & 6.4 & 4.0 & $<$1.12 \\
MJV02916 & 00:14:35.263 & $-$00:13:49.55 & 2.7	 & 5.1	 & 16.5 & 6.2 & 6.4 & $<$1.30 \\
MJV02034 & 00:14:40.426 & $-$00:02:42.87 & 1.3	 & 1.1	 & 16.2 & 6.4 & 2.2 & $<$1.09 \\
MJV02033 & 00:14:44.051 & $-$00:00:18.00 & 1.1	 & 0.8	 & 16.0 & 6.3 & 1.0 & $<$0.99 \\
MJV02904 & 00:14:48.244 & $-$00:27:34.50 & 1.2	 & 1.1	 & 16.3 & 6.5 & 3.3 & $<$0.97 \\
MJV02898 & 00:14:48.260 & $-$00:19:29.94 & 0.8	 & 1.7	 & 16.3 & 6.4 & 2.3 & $<$0.98 \\
MJV02893 & 00:14:48.455 & $-$00:12:49.32 & 10.0	 & 10.5	 & 16.3 & 6.2 & 3.7 & $<$1.30 \\
MJV02901 & 00:14:49.480 & $-$00:27:10.07 & 1.7	 & 1.3	 & 16.3 & 6.5 & 3.0 & $<$0.94 \\
MJV02022 & 00:14:52.452 &   +00:07:00.18 & 0.8	 & 0.9	 & 16.1 & 6.0 & 1.9 & 1.39 \\
MJV02912 & 00:14:54.918 & $-$00:09:27.51 & 6.5	 & 8.0	 & 16.2 & 6.3 & 1.8 & $<$0.88 \\
MJV02911 & 00:14:54.921 & $-$00:09:12.78 & 2.7	 & 3.6	 & 16.2 & 6.3 & 1.8 & $<$0.87 \\
MJV02908 & 00:14:56.210 & $-$00:35:17.32 & 0.9	 & 0.8	 & 16.5 & 6.2 & 5.4 & $<$1.03 \\
MJV02025 & 00:15:00.348 &   +00:05:21.66 & 0.8	 & 0.8	 & 16.0 & 6.1 & 0.3 & $<$1.00 \\
MJV02031 & 00:15:00.852 &   +00:00:48.34 & 0.8	 & 1.6	 & 16.0 & 6.2 & $-$0.6 & $<$0.81 \\
MJV02026 & 00:15:03.275 &   +00:03:18.41 & 2.0	 & 1.7	 & 16.0 & 6.1 & $-$0.4 & $<$0.90 \\
MJV02907 & 00:15:03.402 & $-$00:34:59.36 & 139.9 & 142.6 & 16.4 & 6.2 & 4.6 & 16.73 \\
MJV02042 & 00:15:07.017 & $-$00:08:01.31 & 11.5	 & 12.5	 & 16.4 & 6.2 & $-$0.1 & 2.19 \\
MJV02035 & 00:15:13.988 & $-$00:03:38.14 & 4.3	 & 4.4	 & 16.1 & 6.2 & $-$1.0 & 0.93 \\

\enddata
\tablecomments{
The full (machine-readable) table is available in the online version of this publication.  In the print edition, only the first 20 entries with the first 9 columns are shown.  In the full table, additional columns are available for sources detected in the VLBI observations, including position, integrated flux density, approximate deconvolved size, and the ratio of VLBI--scale flux density to FIRST flux density.
}
\label{tab:catalog}
\end{deluxetable}

\subsection{Browsable catalog}
In addition to the machine--readable table version of the catalog (Table~\ref{tab:catalog}), a html version of the catalog is available at \verb+http://safe.nrao.edu/vlba/mjivs/products.html+.  The  html version of the catalog includes links to additional information for each of the VLBI detections, including contour plots in jpg format and calibrated uv data.

\section{Preliminary results}
\label{sec:results}

\subsection{Detection fractions}
\label{sec:detectionfractions}

In total, 4,336 FIRST sources have been detected by \mjive\ at the time of publication, a number which will continue to grow as observations are added.  21,396 FIRST sources have been imaged at milliarcsecond resolution (excluding the VLBI calibrator sources to avoid the obvious selection bias; however, images for these sources are also available).  This represents a sample around two orders of magnitude larger than previous unbiased VLBI surveys, both imaging \citep[e.g.,][35 detections]{garrington99a} and non--imaging \citep[][85 detections from their preliminary pointing--centre sample]{porcas04a}.  Comparing results with these previous surveys is possible by extracting appropriate subsamples from the \mjive\ survey.  \citet{garrington99a} targeted only FIRST sources with peak flux density $>$10 mJy and largest angular scale $<$5\arcsec, and obtained a 35\% detection fraction with typical detection sensitivity of 1--2 mJy.  By selecting all \mjive\ sources with peak FIRST flux $>$10 mJy and a predicted detection threshold between 1 and 2 mJy, we obtain a subsample of 1283 sources and a detection fraction of 36\%, consistent with \citet{garrington99a}.  \citet{porcas04a} made no pre--selection based on FIRST parameters and had a typical detection sensitivity of 1.1 mJy, with a preliminary analysis yielding a detection fraction of 33\%. By selecting all \mjive\ sources where our predicted detection threshold was between 1.05 and 1.15 mJy, we obtain a subsample of 1466 sources and a detection fraction of 23\%, somewhat lower than \citep{porcas04a}.  This discrepancy is significant (binomial statistics give an error of around 3\% on the \citealt{porcas04a} detection fraction) but the cause is not obvious.  However, the analysis of \citet{porcas04a} is described as ``preliminary", and so one potential explanation is false positives due to noise or confusion, which are difficult to identify in a single--baseline non--imaging observation.

In order to investigate possible evolution of the compact fraction of arcsecond scale sources with decreasing flux density, we have binned our results by FIRST peak flux density and by the ratio of VLBI peak flux density to FIRST peak flux density (hereafter the ``compactness ratio").  Peak \mjive\ flux density was used in preference to integrated flux density because the errors are typically considerably smaller for the former.  The $uv$ coverage and hence beam size and shape can vary between \mjive\ epochs, which affects the measured peak flux density, but a similar effect would also be present in the integrated flux density measurement.  In calculating the detection fraction for each bin, we first exclude all sources where there is a possibility that the source fulfils the bin criteria but would be missed in \mjive\ due to insufficient sensitivity.  Specifically, we loop over all sources with FIRST peak flux densities in the correct range, but exclude any where the FIRST peak flux density multiplied by the minimum compactness ratio for the bin falls below the predicted detection threshold (equal to the predicted image rms obtained from the pipeline multiplied by our minimum detection threshold of 6.75).  The remaining sample which passes all tests is the valid sample for that bin.  The number of detections from the valid sample is calculated and divided by the size of the valid sample for the bin to obtain the detection fraction for that bin.   

Theoretically, this approach excludes all fields where there is a chance that a valid source could go undetected.  In practice, several effects may alter the detectability of a source in any individual field very near the threshold, such as the scatter between predicted and actual image rms and source variability.  However, since both of these effects introduce a relatively low degree of scatter and are essentially zero--mean, and furthermore there are no drastic changes seen between bins, we neglect them in the following analysis.

\begin{figure}
\begin{center}
\includegraphics[width=0.85\textwidth]{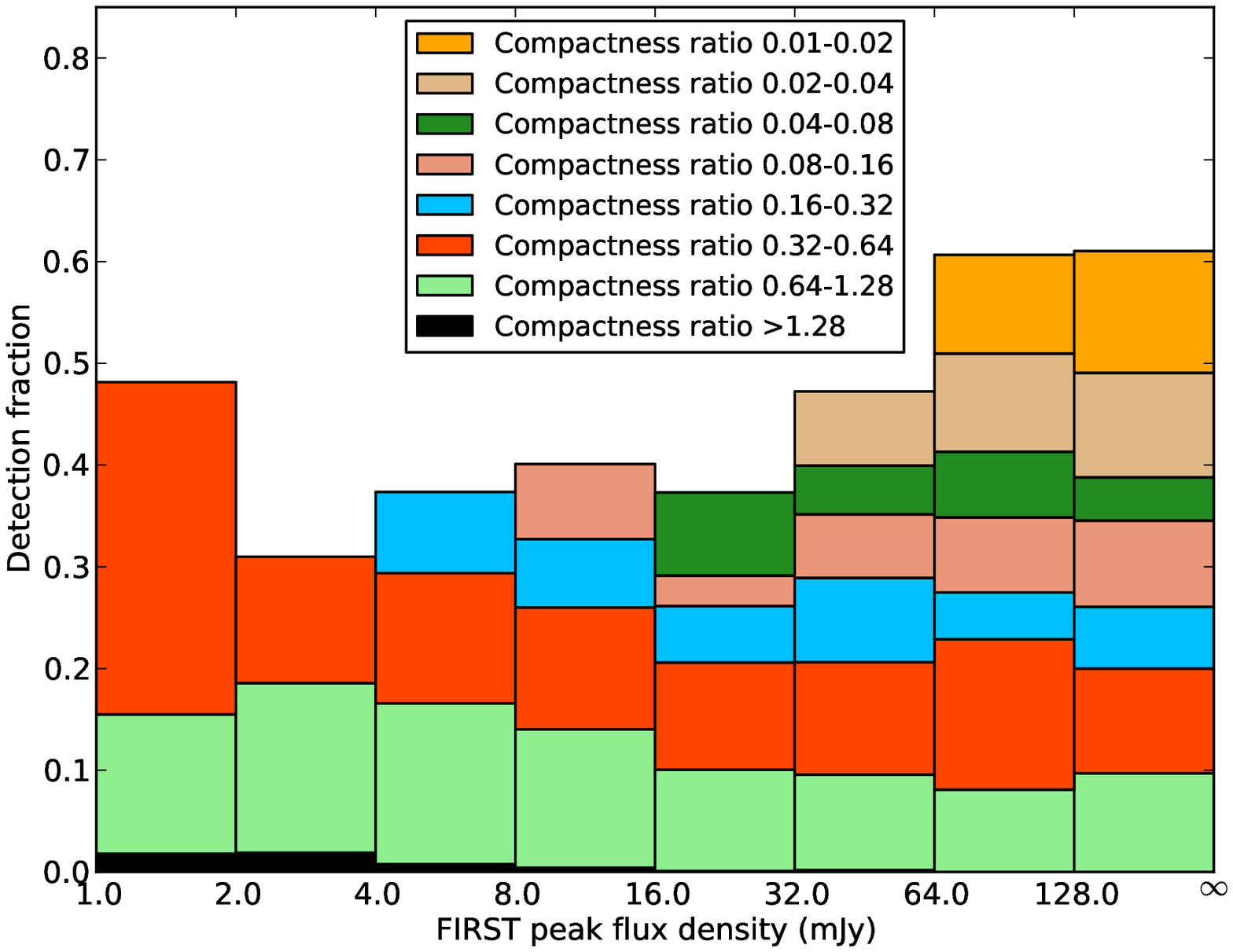}
\caption{Detection fraction for \mjive\ sources, binned by FIRST flux.  The different
colors show the detection fraction when probing down to a progressively smaller
compactness ratio (ratio of peak VLBI flux density to peak FIRST flux density). Very small
compactness ratios can only be probed by the brighter sources in the \mjive\ sample, and
so the smallest compactness ratio are probed in progressively fewer bins. 
For sources exceeding a given compactness
ratio, it can be seen that the detection fraction generally rises slightly for fainter
FIRST sources.  Only bins with more than 50 sources are shown in this plot. 
Full details of all bins including error bars are shown in Table~\ref{tab:detections}.}
\label{fig:detections}
\end{center}
\end{figure}

The binned cumulative detection fractions, and their error bars, are presented in Table~\ref{tab:detections}. 
The lower left quadrant of Table~\ref{tab:detections} is unsampled, because it  corresponds to faint FIRST sources with only a small component of the flux remaining compact, and these sources would fall below our detection limits.  The detection fractions are also shown in a stacked column format in Figure~\ref{fig:detections}.  In Figure~\ref{fig:detections}, bins with fewer than 50 target fields meeting the criteria are not shown, to avoid confusing the plot with results with large errors. 

Figure~\ref{fig:detections} and Table~\ref{tab:detections} reveal a trend towards an increased likelihood of a compact component for sources with a lower arcsecond--scale radio flux density.  The difference is statistically significant across the range of compact fractions for which we have adequate data, and is insensitive to changes in number of FIRST flux density bins used.  At both very low and very high compact fractions, an insufficient number of sources are available; at low compact fractions this is due to an inability to probe the faint end of the flux density distribution, while at high compact fractions the intrinsic detection rate is low.  Accordingly, we focus on the bin for compact fractions $>$0.32, which has the highest total number of \mjive\ detections (and the most even distribution across the whole range of arcsecond--scale flux densities).  Similar results are seen for compactness ratios $>$0.64 and $>$0.16.  A simple weighted linear regression to the data presented in Table~\ref{tab:detections} shows that our results are inconsistent with no evolution of compactness with arcsecond--scale flux density.  The reduced $\chi^2$ of a model in which compactness was independent of arcsecond--scale radio flux (hereafter the ``reference" model) was 8.8, whereas for the weighted linear regression fit, the reduced $\chi^2$ was 2.2.  The best linear fit was obtained when detection fraction increased by 0.025 with each halving of arcsecond--scale flux density.  The relatively poor fit can be attributed to the fact that a simple linear model covering the entire \mjive\ sample is likely  inadequate -- this is explored further in Section~\ref{sec:sdssassoc}.

Across the range of fluxes sampled by \mjive, the source population is expected to be dominated by radio loud AGN, as the well--known shoulder seen in radio source counts does not appear until flux densities $\lesssim$1 mJy \citep[e.g.,][]{hopkins03a,norris11a}.  Thus, an evolution in the compactness of the \mjive\ sources was not necessarily an expected result.  Previously, an anti--correlation between total flux density and VLBI compactness has been noted in observations at higher frequencies \citep{lawrence85a} when studying bright ($\gtrsim$1 Jy) sources, but this falls well outside the \mjive\ flux range.  

A possible explanation for the anti--correlation between arcsecond--scale flux density and VLBI compactness lies in the kinematics of the AGN radio jet.  \citet{mullin08a} show a significant anti--correlation between luminosity and core prominence for broad line radio galaxies and narrow line radio galaxies, which they attribute to Doppler boosting.  Since higher luminosity sources have on average a higher jet bulk Lorentz factor $\Gamma$, and the solid angle within which Doppler boosting occurs (where the Doppler factor $\delta > 1$) becomes smaller as $\Gamma$ increases, a higher luminosity source is less likely to be Doppler boosted and more likely to be suppressed when seen from an arbitrary viewing angle.  Conversely, lower--power radio sources, with slower jets, are less likely to be Doppler suppressed and hence a greater fraction of the arcsecond--scale radio flux is likely to come from a compact core.  Since the Doppler boosting depends sensitively on the viewing angle, significant differences could be expected depending on the source classification.  This is investigated in Section~\ref{sec:sdssassoc}.

Many other effects could also influence the number of observed sources with a given compact fraction at a given flux density to a greater or lesser degree.  For instance, $\delta$ affects the total apparent radio luminosity and hence the observable volume for a flux--density--limited sample, so while sources with large $\delta$ (and hence typically a high compact fraction) are less common in a given volume, they are also observable out to larger distances. The relative contribution of extended radio lobe emission to the total source flux density may also change as a function of flux density.  A detailed analysis synthesizing the impact of all possible factors is beyond the scope of this paper.  In a future paper, we will perform a more sophisticated analysis of the distribution of compact fractions, and will combine the \mjive\ results after a further expansion of the sample size with sub--mJy sources from deep fields \citep[e.g.,][]{middelberg13a} and brighter VLBI sources identified from calibrator surveys \citep[e.g.,][]{beasley02a} to investigate the properties of the compact radio source population across 5 orders of magnitude in flux density.
 
\begin{deluxetable}{lrrrrrrrr}
\rotate
\tabletypesize{\scriptsize}
\addtolength{\tabcolsep}{-3.5pt}
\tablecaption{Detection fractions as a function of FIRST brightness and VLBI compactness. Number of potential sources in each bin is shown in parentheses.}
\tablewidth{0pt}
\tablehead{
\colhead{Compact} & \colhead{FIRST flux} & \colhead{FIRST flux} & \colhead{FIRST flux} & \colhead{FIRST flux} & \colhead{FIRST flux} & \colhead{FIRST flux} & \colhead{FIRST flux} & \colhead{FIRST flux} \\\colhead{flux ratio} & \colhead{1 - 2 mJy} & \colhead{2 - 4 mJy} & \colhead{4 - 8 mJy} & \colhead{8 - 16 mJy} & \colhead{16 - 32 mJy} & \colhead{32 - 64 mJy} & \colhead{64 - 128 mJy} & \colhead{$>$ 128 mJy}}
\startdata
$>$ 1.28 & $0.018^{+0.002}_{-0.002}$ (7243) & $0.019^{+0.002}_{-0.002}$ (4593) & $0.008^{+0.002}_{-0.002}$ (2697) & $0.004^{+0.002}_{-0.002}$ (1677) & $0.001^{+0.002}_{-0.001}$ (1015) & $0.002^{+0.004}_{-0.002}$ (543) & $0.000^{+0.006}_{-0.000}$ (284) & $0.000^{+0.011}_{-0.000}$ (165)  \\
$>$ 0.64 & $0.155^{+0.008}_{-0.007}$ (2345) & $0.186^{+0.007}_{-0.007}$ (3371) & $0.166^{+0.007}_{-0.007}$ (2598) & $0.140^{+0.009}_{-0.008}$ (1677) & $0.100^{+0.010}_{-0.009}$ (1015) & $0.096^{+0.014}_{-0.012}$ (543) & $0.081^{+0.019}_{-0.016}$ (284) & $0.097^{+0.028}_{-0.023}$ (165)  \\
$>$ 0.32 & $0.481^{+0.075}_{-0.074}$ \phn\phn(54) & $0.310^{+0.014}_{-0.014}$ (1152) & $0.294^{+0.011}_{-0.010}$ (1913) & $0.260^{+0.011}_{-0.011}$ (1623) & $0.206^{+0.013}_{-0.013}$ (1015) & $0.206^{+0.018}_{-0.017}$ (543) & $0.229^{+0.027}_{-0.025}$ (284) & $0.200^{+0.035}_{-0.031}$ (165)  \\
$>$ 0.16 & $-$ \phn\phn\phn(0) & $0.323^{+0.103}_{-0.090}$ \phn\phn(31) & $0.374^{+0.020}_{-0.020}$ \phn(613) & $0.327^{+0.014}_{-0.013}$ (1201) & $0.261^{+0.014}_{-0.014}$ \phn(983) & $0.289^{+0.020}_{-0.019}$ (543) & $0.275^{+0.028}_{-0.027}$ (284) & $0.261^{+0.038}_{-0.035}$ (165)  \\
$>$ 0.08 & $-$ \phn\phn\phn(0) & $-$ \phn\phn\phn(0) & $0.462^{+0.169}_{-0.162}$ \phn\phn(13) & $0.401^{+0.025}_{-0.025}$ \phn(404) & $0.291^{+0.017}_{-0.017}$ \phn(731) & $0.352^{+0.022}_{-0.021}$ (512) & $0.349^{+0.030}_{-0.029}$ (284) & $0.345^{+0.040}_{-0.038}$ (165)  \\
$>$ 0.04 & $-$ \phn\phn\phn(0) & $-$ \phn\phn\phn(0) & $-$ \phn\phn\phn(0) & $0.538^{+0.162}_{-0.169}$ \phn\phn(13) & $0.373^{+0.032}_{-0.031}$ \phn(252) & $0.399^{+0.026}_{-0.025}$ (393) & $0.413^{+0.031}_{-0.030}$ (276) & $0.388^{+0.041}_{-0.039}$ (165)  \\
$>$ 0.02 & $-$ \phn\phn\phn(0) & $-$ \phn\phn\phn(0) & $-$ \phn\phn\phn(0) & $-$ \phn\phn\phn(0) & $0.429^{+0.244}_{-0.219}$ \phn\phn\phn(7) & $0.472^{+0.047}_{-0.047}$ (127) & $0.510^{+0.036}_{-0.036}$ (208) & $0.491^{+0.041}_{-0.041}$ (161)  \\
$>$ 0.01 & $-$ \phn\phn\phn(0) & $-$ \phn\phn\phn(0) & $-$ \phn\phn\phn(0) & $-$ \phn\phn\phn(0) & $-$ \phn\phn\phn(0) & $1.000^{+0.000}_{-0.594}$ \phn\phn(2) & $0.607^{+0.067}_{-0.070}$ \phn(61) & $0.610^{+0.043}_{-0.045}$ (136) 
\enddata
\label{tab:detections}
\end{deluxetable}

\subsection{SDSS associations}
\label{sec:sdssassoc}

One of the powerful aspects of the FIRST survey is its overlap with SDSS, allowing identification of optical counterparts to many of the VLBI detections (and non--detections).  We have separated the \mjive\ sample into detections and non--detections to investigate their optical characteristics.  For this purpose, we made use of the classification assigned to the original FIRST objects (which were made using the nearest source within a maximum radius of 8 arcseconds in the DR9 data release; \citealp{ahn12a}).  This has the disadvantage of not taking into account the better positional accuracy obtained from the VLBI detections for detected sources, but allows for a more uniform treatment of non--detections.  Of the 21,396 sources targeted by \mjive\ to date, 40\% are classified as galaxies on the basis of SDSS information (8,493 sources, 1,788 detections), 15\% are classified as stellar/point--like (3,165 sources, 960 detections), and 45\% are unclassified (9,738 sources, 1,558 detections).  In light of the \mjive\ detections, the SDSS classification of ``stellar" is obviously incorrect for virtually all of these sources -- they are distant, compact AGN sources.  However, to maintain consistency with the classification used in the FIRST catalog, we will continue to refer to this source class as stellar/point--like.

\begin{figure}
\begin{center}
\includegraphics[width=\textwidth]{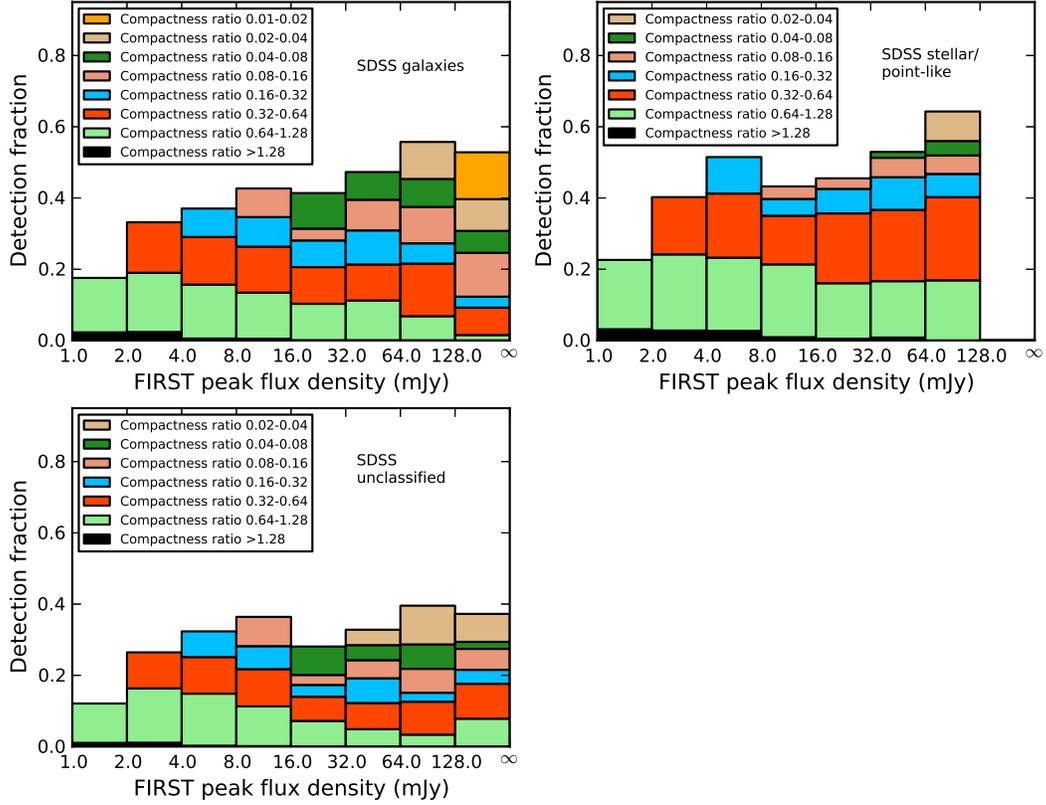}
\caption{Detection fraction for \mjive\ sources, binned by FIRST flux and separated by SDSS classification.  The
same bins, minimum source counts and terminology are used as in Figure~\ref{fig:detections}.
The top left plot shows sources which are classified as galaxies by SDSS, the top right plot 
shows stellar/point-like sources, and the bottom left plot shows sources with no optical identification in SDSS.
Stellar/point-like optical sources (presumably AGN) have a higher probability of detection overall,
but this detection fraction is consistent with no variation with arcsecond--scale radio flux density.  Galaxies and sources which
are undetected in SDSS show much stronger evolution with radio flux density.}
\label{fig:sdssdetections}
\end{center}
\end{figure}

When the \mjive\ sample is separated by SDSS classification as described above, a clear trend emerges, as shown in Figure~\ref{fig:sdssdetections}.  Stellar/point--like SDSS sources are considerably more likely to be detected as VLBI sources than either galaxies or sources which are have no SDSS counterpart. For example, as before considering sources with a compactness fraction $>$0.32, a stellar/point--like SDSS source has a detection probability of around 35 -- 40\%, compared with 10 -- 30\% for other sources (galaxies or unclassified).  

However, the detection fraction of sources classified as stellar/point--like in SDSS appears to be independent of FIRST flux density, whereas the remaining sources (galaxies and unclassified) show a strong anti--correlation between FIRST flux density and VLBI compactness.  The trend for fainter radio sources in these latter two categories to be more compact is even clearer than for the complete \mjive\ dataset (as shown in Figure~\ref{fig:detections}), since a significant population of sources which show no compactness evolution with FIRST flux density (the SDSS stellar/point--like sources) has been removed.  Following the procedure used for the full dataset, we made weighted linear regression fits to the evolution of compact fraction with FIRST peak flux density for each SDSS class separately.  The results are summarized in Table~\ref{tab:sdssfits}, and show that the linear fit to compactness as a function of FIRST flux density is generally better for each separate SDSS class than when the entire \mjive\ sample is considered as a whole.

%This result provides strong support for the unification model proposed in \citet{mullin08a}, where Doppler boosting and suppression explain the variations in core prominence between classes of sources.  Sources with Doppler boosted core emission are more likely to be seen as stellar/point--like objects in SDSS, and this Doppler boosting makes the core emission more prominent on VLBI scales, yielding the higher average VLBI detection rate.  Sources which are not strongly Doppler boosted are more likely to appear as galaxies or non--detections in SDSS, and these sources are less likely to be detected overall.  However, for these sources, a lower luminosity and hence lower $\gamma$ will lead on average more core prominence, with the wider boosting solid angle causing either a transition from de--boosting to boosting, or at least a reduction in the de--boosting.

This result is consistent with the unification model proposed
in \citet{mullin08a}, where Doppler boosting and suppression explain the variations in core prominence between classes of sources. However, the information available in SDSS is not detailed enough to directly place \mjive\ sources into the classification schemes used by \citet{mullin08a}, and therefore a comparison can only be qualitative.

We coarsely identify the stellar/point--like SDSS sources with the categories of quasars and broad line radio galaxies from \citet{mullin08a}.  Regardless of source luminosity, the orientation of most of these sources will mean that the core emission will be Doppler boosted with high Lorentz factors, and they will therefore have relatively high (and constant) detection fractions across the range studied by \mjive.  Supporting this interpretation, other previous studies with small samples have also noted comparatively high levels of radio core prominence in quasars \citep[][]{morganti97a}.

Sources similar to the narrow-line radio galaxies (NLRGs) and low-excitation radio galaxies (LERGs) from Mullin et al. are more likely to be found among the SDSS sources classified as ``galaxies" or the ``unclassified" SDSS objects. In these sources, the viewing angles are spread over a wider range, but as the Lorentz factors are smaller, the effects of boosting and de--boosting are less pronounced. In these source classes, lower-luminosity sources have on average a higher core prominence than higher--luminosity sources, due to reduced Doppler suppression of the core emission. Qualitatively, this  leads to the observed situation in which SDSS galaxies and unclassified SDSS objects have higher detection fractions at lower flux densities.

%Figures~\ref{fig:detections} and \ref{fig:sdssdetections} shows intriguing evidence that the detection fraction begins to turn over in the range 1 -- 2 mJy for sources classified as galaxies or non--detections, precisely in the region where star forming galaxies begin to make a contribution to the radio source counts.  Although based on a single bin (the 1 -- 2 mJy FIRST sources, with a compactness ratio $>$ 0.64), the reduction in detection fraction is quite marked.  However, drawing conclusions based on this single data point would be premature, and much stronger evidence will be possible when the \mjive\ sample is combined with deeper VLBI survey data, such as that in \citet{middelberg13a} and similar (but larger and deeper) studies of the COSMOS field which are now underway.  These deeper surveys will both boost the number counts and allow the $\sim$mJy sources to be probed to fainter compactness ratios.
%(and w, core emission will be doppler boosted, while for most narrow line sources the core emission will be de--boosted.  However, for low--luminosity sources with a lower $\gamma$, the effects of boosting and de--boosting will be less pronounced, leading to a relatively higher core prominence for the lower luminosity narrow line sources.  Since most narrow line radio galaxies will be seen as optical galaxies or non--detections in SDSS, the observed trends in compactness ratio can be expected to emerge.

\begin{deluxetable}{llll}
\tabletypesize{\scriptsize}
\tablecaption{Compact fraction evolution, separated by SDSS type}
\tablewidth{0pt}
\tablehead{
\colhead{SDSS type} & \colhead{Best--fit linear gradient} & \colhead{Linear fit reduced $\chi^2$} & \colhead{Reference model reduced $\chi^2$}}
\startdata
Combined \mjive\ dataset & -0.025 & 2.22 & 8.82 \\
Galaxy & -0.043 & 2.66 & 8.18 \\
Stellar/point--like & -0.009 & 1.96 & 1.85 \\
Undetected & -0.043 & 1.66 & 8.88 
\enddata
\label{tab:sdssfits}
\end{deluxetable}

\subsection{Highly variable sources}
\label{sec:variable}
Simply due to light travel time arguments, radio sources varying on a timescale of years must contain a reasonably compact component.  They therefore form a particularly interesting subclass of objects when considering a VLBI survey such as \mjive, which is sensitive only to compact sources.  Examples of known variable sources which might be expected to be detected include blazars, x--ray binaries, scintillating radio pulsars or AGN, gamma--ray burst afterglows, radio supernovae, and supernova remnants, whilst previously unsuspected explosive phenomena could also be found if present.  A number of surveys for slow variability are planned with upcoming telescopes \citep[e.g., ASKAP, LOFAR;][]{johnston08a,stappers11a}, but none of these will be able to provide information on milliarcsecond scale structure as is available with the VLBA.  Identifying and classifying variable radio sources with \mjive, then, provides a useful complement to previous and future studies of slow radio transients.

Previous studies \citep[e.g.,][]{levinson02a,de-vries04a,bower07a} have identified variable sources by using the publicly available data from the FIRST survey, sometimes in combination with other surveys such as the NRAO VLA Sky Survey \citep[NVSS;][]{condon98a}. The largest of these, conducted by \citet{thyagarajan11a}, compared sources detected in at least one observation from FIRST and/or NVSS and identified 1627 variable (detected in multiple epochs) and transient (detected in a single epoch, with inconsistent upper limits in other epochs) sources.  Of the 1627 sources in the \citet{thyagarajan11a} catalog, 19 have been observed by \mjive\ to date, and 8 were detected.  All of the observed sources are listed in Table~\ref{tab:knownvariables}.  As expected, the \mjive\ detection fraction amongst confirmed variable sources is higher than for the general radio source population. 
%Two of the non--detected sources (MJV09296 and MJV18927) are flagged as a possible sidelobe in the \citet{thyagarajan11a} catalog because they are near another bright radio source; if these are excluded, the \mjive\ detection fraction is 50\%.  As expected, this detection fraction amongst confirmed variable sources is considerably higher than for the general radio source population.  The non--detection of the bright  MJV07763 and MJV19059 and 
%It is also notable that for all of the sources in Table~\ref{tab:knownvariables}, the VLBI flux density is less than or equal to the FIRST flux density (to within the errors).  This illustrates that variable sources are most likely to be discovered when in a bright state 

\begin{deluxetable}{lllll}
\tabletypesize{\tiny}
\tablecaption{Known variable sources observed by \mjive}
\tablewidth{0pt}
\tablehead{
\colhead{FIRST} & \colhead{\mjive} & \colhead{FIRST flux density} & \colhead{\mjive\ flux density} & \colhead{Flux} \\\colhead{identifier} & \colhead{identifier} & \colhead{(mJy/beam)} & \colhead{(mJy/beam)} & \colhead{ratio}}
\startdata
FIRST J160404.237+313925.09 & MJV11238 & 1.2 & $<$ 0.9 & $<$ 0.77  \\
FIRST J075307.619+483429.74 & MJV11698 & 1.3 & $<$ 2.1 & $<$ 1.63  \\
FIRST J133715.322-052837.55 & MJV10634 & 3.0 & $<$ 0.6 & $<$ 0.21  \\
FIRST J135717.919+192243.15 & MJV10688 & 11.5 & 6.8 & 0.59  \\
FIRST J161902.527+225355.11 & MJV03614 & 1.7 & 1.3 & 0.77  \\
FIRST J094507.299+101458.08 & MJV09783 & 2.8 & 3.0 & 1.06  \\
FIRST J151726.334+200116.72 & MJV11088 & 2.9 & $<$ 1.2 & $<$ 0.42  \\
FIRST J074754.818+313436.75 & MJV18927 & 3.9 & $<$ 2.4 & $<$ 0.61  \\
FIRST J080706.323+443836.93 & MJV19059 & 42.3 & $<$ 2.3 & $<$ 0.06  \\
FIRST J164042.063+394825.71 & MJV09296 & 17.1 & $<$ 0.9 & $<$ 0.05  \\
FIRST J080623.978+444247.12 & MJV18996 & 3.5 & 1.9 & 0.54  \\
FIRST J150605.109+373755.67 & MJV07763 & 9.0 & $<$ 0.6 & $<$ 0.07  \\
FIRST J150456.228+095919.03 & MJV09187 & 1.4 & $<$ 2.2 & $<$ 1.57  \\
FIRST J114315.209+161809.35 & MJV22015 & 4.6 & $<$ 4.4 & $<$ 0.95  \\
FIRST J150757.784+423343.60 & MJV04830 & 3.9 & 2.6 & 0.66  \\
FIRST J085418.962+200346.61 & MJV01666 & 16.1 & 1.5 & 0.09  \\
FIRST J154242.609+145957.76 & MJV07897 & 5.7 & 3.8 & 0.66  \\
FIRST J131654.888+410508.69 & MJV16238 & 1.9 & $<$ 0.7 & $<$ 0.35  \\
FIRST J073826.371+294628.92 & MJV13634 & 38.9 & 20.6 & 0.53 
\enddata
\label{tab:knownvariables}
\end{deluxetable}

In addition to investigating the VLBI characteristics of previously identified variable sources, it is possible to cross--match the \mjive\ and FIRST fluxes to identify previously unknown variable sources. The mismatched resolution between FIRST and \mjive\ means that it is impossible to identify any sources which have decreased in flux since the FIRST observations, since this situation is degenerate with the much more common case of some flux being resolved out on intermediate scales.  It is, however, possible to identify some sources which have increased significantly in flux.  

The accuracy of peak fluxes for objects detected in the FIRST survey are thought to range between $\sim$15\% at the detection limit (where the errors are dominated by the thermal noise in the maps) to 5\% or less for brighter sources \citep{white97a}.  Since most of the detected variable sources are faint, we allow a 15\% error for the FIRST flux density irrespective of brightness.  For the \mjive\ detections, we first conservatively allow for a 20\% error in the absolute flux density calibration, and then calculate the 3$\sigma$ confidence lower limit of the adjusted peak flux density based on the signal--to--noise of the detection.  Only sources where this \mjive\ lower limit exceeds the FIRST upper limit are considered to be confirmed variable sources.

In the \mjive\ sample of 4,336 detections, approximately 1\% exhibit significant variability, with the most variable source increasing in flux by a factor of more than 3.  Due to the selection effects mentioned above, 1\% is clearly a lower limit on the number of variable sources in the sample (as probed over a timescale of $\sim$10 years).  These sources are shown in Table~\ref{tab:newvariables}.  

\begin{deluxetable}{lllll}
\tabletypesize{\tiny}
\tablecaption{Variable sources discovered by \mjive}
\tablewidth{0pt}
\tablehead{
\colhead{FIRST} & \colhead{\mjive} & \colhead{FIRST flux density} & \colhead{\mjive\ flux density} & \colhead{Flux} \\\colhead{identifier} & \colhead{identifier} & \colhead{(mJy/beam)} & \colhead{(mJy/beam)} & \colhead{ratio}}
\startdata
MJV03427 & FIRST J132752.087+221623.18 & 6.3 & 10.9 & 1.7  \\
MJV09065 & FIRST J144515.076+173005.38 & 1.2 & 2.5 & 2.1  \\
MJV12257 & FIRST J103941.807+052128.50 & 2.5 & 4.5 & 1.8  \\
MJV05862 & FIRST J124212.348+372200.44 & 1.9 & 3.7 & 2.0  \\
MJV21296 & FIRST J230008.056+033607.54 & 1.3 & 2.4 & 1.8  \\
MJV20585 & FIRST J125856.917+143743.10 & 4.6 & 9.0 & 2.0  \\
MJV02969 & FIRST J010848.869+014056.98 & 1.1 & 3.3 & 3.0  \\
MJV19115 & FIRST J082316.467+621209.76 & 2.9 & 7.5 & 2.6  \\
MJV02676 & FIRST J233437.938+072626.98 & 1.6 & 3.7 & 2.3  \\
MJV02677 & FIRST J233243.694+072423.16 & 1.1 & 2.5 & 2.3  \\
MJV07228 & FIRST J122131.360+442759.63 & 2.2 & 4.8 & 2.2  \\
MJV15170 & FIRST J224601.781+034820.01 & 1.5 & 2.9 & 1.9  \\
MJV10933 & FIRST J144101.810+381055.01 & 2.3 & 3.7 & 1.6  \\
MJV09125 & FIRST J150404.221+102253.65 & 2.2 & 4.1 & 1.8  \\
MJV16559 & FIRST J162850.575+222827.11 & 1.3 & 3.0 & 2.3  \\
MJV11861 & FIRST J095745.499+245644.79 & 15.3 & 37.7 & 2.5  \\
MJV15680 & FIRST J112100.831+180557.11 & 2.0 & 4.2 & 2.1  \\
MJV05224 & FIRST J082650.527+354929.43 & 1.4 & 2.5 & 1.8  \\
MJV16621 & FIRST J080942.982+574232.25 & 2.0 & 4.8 & 2.4  \\
MJV06379 & FIRST J082401.275+560052.63 & 7.2 & 11.1 & 1.5  \\
MJV04039 & FIRST J081337.037+364314.44 & 1.4 & 3.0 & 2.1  \\
MJV15116 & FIRST J224520.019+031350.18 & 3.2 & 5.3 & 1.6  \\
MJV07956 & FIRST J160315.164+330956.87 & 2.3 & 4.5 & 2.0  \\
MJV11709 & FIRST J075143.060+331255.78 & 2.3 & 6.1 & 2.7  \\
MJV11032 & FIRST J151643.700+192104.98 & 1.8 & 3.0 & 1.7  \\
MJV10782 & FIRST J142719.935+541720.09 & 1.9 & 3.6 & 1.9  \\
MJV16337 & FIRST J131323.310+525518.88 & 1.3 & 2.9 & 2.2  \\
MJV21805 & FIRST J123134.285+245500.57 & 3.5 & 5.7 & 1.6  \\
MJV08650 & FIRST J124007.847+050704.81 & 1.9 & 3.5 & 1.9  \\
MJV16982 & FIRST J132644.980+434804.65 & 7.7 & 12.5 & 1.6  \\
MJV03775 & FIRST J222704.241+004517.46 & 6.8 & 12.8 & 1.9  \\
MJV07765 & FIRST J150646.349+373740.16 & 8.9 & 14.7 & 1.6  \\
MJV01680 & FIRST J085454.868+194558.35 & 1.1 & 2.3 & 2.1  \\
MJV17736 & FIRST J130521.336+495142.28 & 10.5 & 20.5 & 2.0  \\
MJV16532 & FIRST J162629.782+230253.90 & 3.2 & 6.5 & 2.0  \\
MJV11563 & FIRST J015511.304+001550.50 & 0.8 & 3.5 & 4.4  \\
MJV04487 & FIRST J115635.486+165310.79 & 2.4 & 4.3 & 1.8  \\
MJV02880 & FIRST J131159.374+322158.19 & 3.4 & 5.7 & 1.7  \\
MJV13841 & FIRST J002941.192+054744.23 & 1.9 & 3.9 & 2.1  \\
MJV01591 & FIRST J080821.489+495253.51 & 1.5 & 3.2 & 2.1  \\
MJV11394 & FIRST J165729.544+481630.08 & 3.6 & 7.3 & 2.0  \\
MJV17137 & FIRST J080637.069+510921.19 & 1.3 & 2.5 & 1.9  \\
MJV16269 & FIRST J131708.713+413103.49 & 2.7 & 4.7 & 1.7  \\
MJV06368 & FIRST J082314.694+560949.02 & 6.4 & 11.9 & 1.9  \\
MJV09587 & FIRST J083227.716+240309.69 & 1.3 & 2.7 & 2.1  \\
MJV07340 & FIRST J125437.693+114304.56 & 1.2 & 2.3 & 1.9  \\
MJV10065 & FIRST J111545.315+081459.36 & 1.6 & 5.1 & 3.2  \\
MJV14899 & FIRST J163652.297+382950.90 & 1.6 & 4.4 & 2.7  \\
MJV09327 & FIRST J164154.244+400032.00 & 6.9 & 15.1 & 2.2  \\
MJV13585 & FIRST J221914.298+015645.52 & 1.2 & 3.4 & 2.8  \\
MJV07914 & FIRST J160213.440+331140.29 & 3.3 & 5.2 & 1.6  \\
MJV16709 & FIRST J111825.991+602929.43 & 2.4 & 5.8 & 2.4  \\
MJV15158 & FIRST J224642.026+030312.51 & 1.9 & 3.8 & 2.0 
\enddata
\label{tab:newvariables}
\end{deluxetable}

\subsection{Compact double sources}
\label{sec:doubles}
Compact Symmetric Objects (CSO) are radio sources structurally
reminiscent of Fanaroff-Riley type II sources, but are several
orders of magnitudes smaller in diameter ($<1$\,kpc), and so are
typically contained within their host galaxies. They are related
to the Gigahertz-Peaked Spectrum sources (GPS) and Compact
Steep-Spectrum sources (CSS) in that many GPS and CSS objects
exhibit CSO morphologies, a fact arising from synchrotron
self-absorption \citep{de-vries09a}. They are thought to be
young objects with ages $<10^4$\,yr \citep{readhead96a},
which eventually may evolve into FR\,II sources. Their ages can
be derived geometrically from the increasing separation of their
lobes, and statistics show that many CSOs are even younger than
500\,yr \citep{gugliucci05a}. This finding implies that CSOs
are short-lived sources. The angle between the jet axis and the
line of sight is typically large in CSOs, and so their jets are
not strongly beamed, which makes them useful probes for aspects
of the unification scenario of AGN \citep{antonucci93a},
when parts of the circumnuclear material can be viewed in
absorption against the receding lobes or jets. Since they are so
small, observations of CSOs are mostly limited to VLBI
observations, which unfortunately implied until recently that
only small samples with order 10--20 objects could be
observed. The largest sample of CSOs observed to date is the
sample by \citet{tremblay09a}, who isolated 103 CSOs from a
parent sample of 1127 flat-spectrum radio sources observed with
the VLBA at 5\,GHz \citep{helmboldt07a}. Our survey will
boost the number of known CSOs simply because it targets so many
objects.

\begin{figure}
\begin{center}
\includegraphics[width=0.5\textwidth, angle=270]{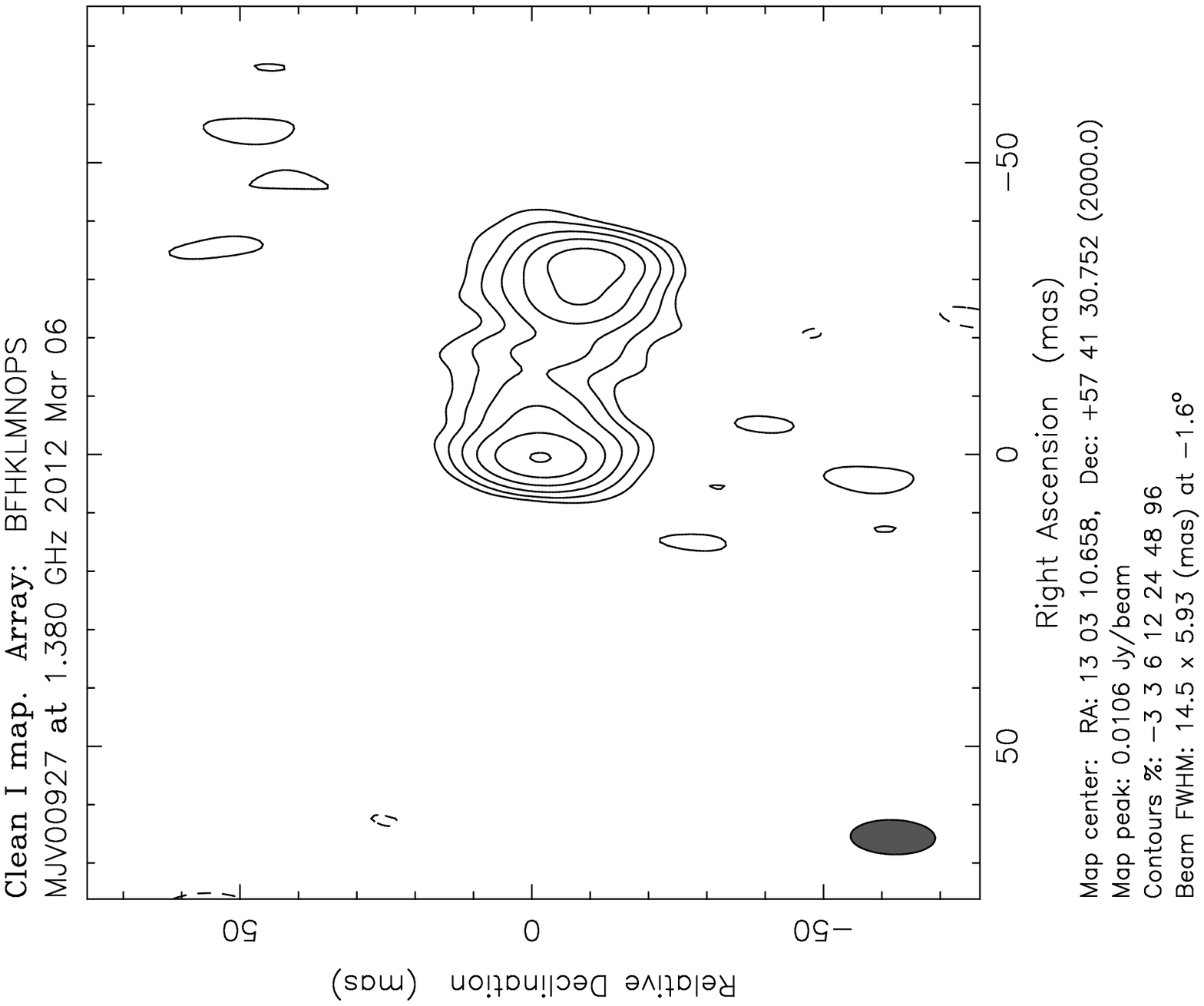}
\includegraphics[width=0.5\textwidth, angle=270]{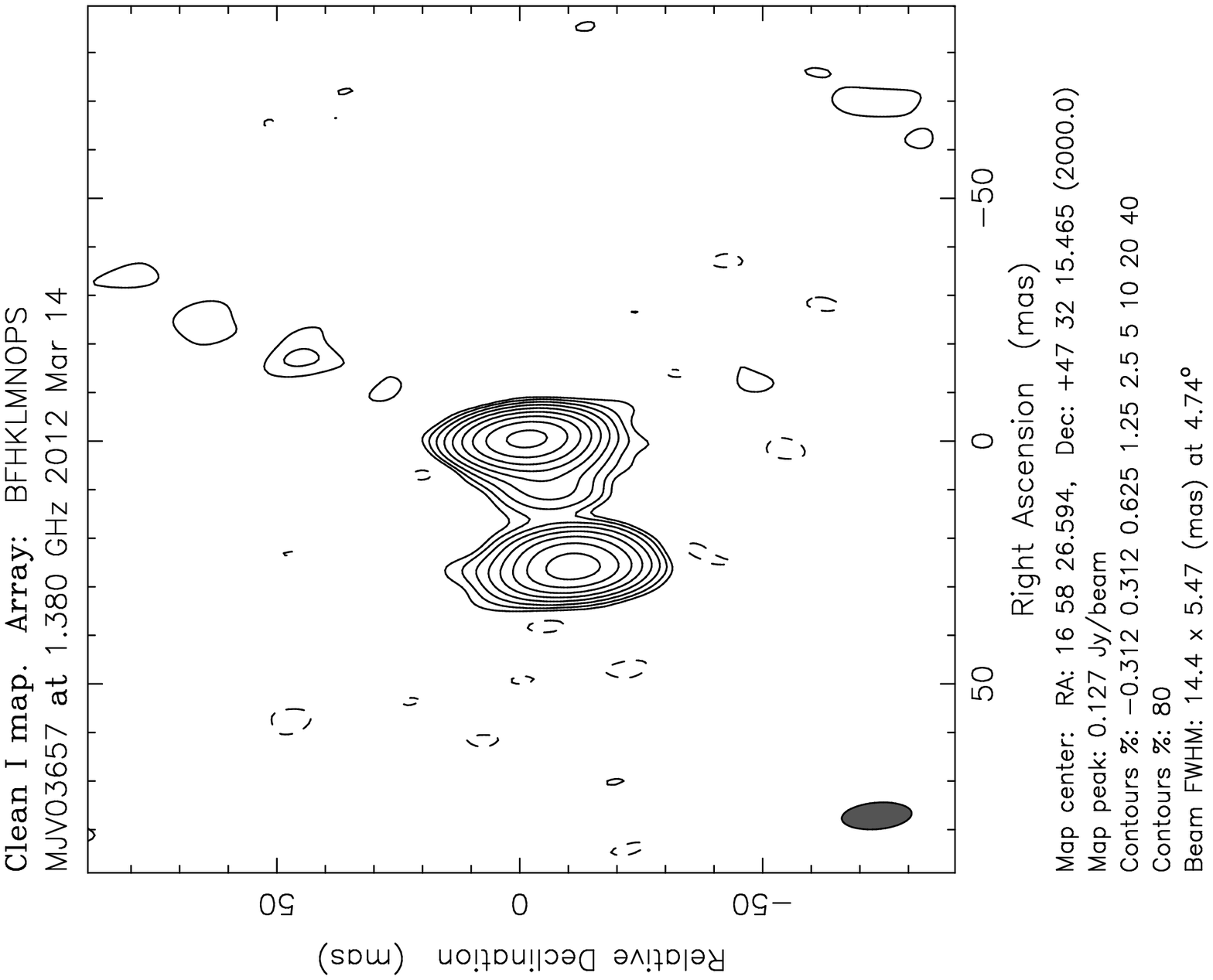}
\caption{Two examples of compact double sources identified by \mjive.  {\em (Left)} MJV00927/FIRST J130310.6+574130, {\em (Right)} MJV03657/FIRST 165826.5+473215.}
\label{fig:compactdoubles}
\end{center}
\end{figure}

A significant number of compact double sources have already been observed by \mjive.  Two examples are shown in Figure~\ref{fig:compactdoubles}. However, reliably identifying compact double sources automatically in the \mjive\ catalog is challenging (the two images shown in Figure~\ref{fig:compactdoubles} were generated manually after a cursory inspection of the automated results by eye).  In particular, allowing the cleaning step to locate flux over a wider field runs the risk of distorting the properties of simpler sources, by placing clean components in sidelobes in the dirty map.  This problem is particularly acute for \mjive\ due to the high sidelobe level in the snapshot VLBI observations.  Accordingly, we plan to implement a separate imaging step which will be focused solely on identifying extended and double sources, and which will supplement rather than replace the main \mjive\ imaging pipeline.  Options under consideration include clean auto--boxing and model--fitting.  We plan to describe the results of this analysis in a future paper.

\section{Conclusions}
\label{sec:conclusions}

The \mjive\ large VLBA project has detected 4,336 compact radio sources in 364 hours of observing time to date.  The overall detection fraction of VLBI sources is consistent with that seen in previous, much smaller imaging surveys.  Using the \mjive\ catalog, we have shown for the first time that the fraction of mJy--level radio sources which contain a compact component has a dependence on the arcsecond--scale radio flux density.  When separated by optical classification using SDSS, further trends emerge, with point--like objects identified as stellar sources by SDSS (and hence presumably compact optical AGN in actuality) being more likely to be detected in VLBI observations overall, but showing no dependence on the arcsecond--scale flux density.  Galaxies and optically undetected sources are less likely to show a VLBI component on average, but this likelihood is a function of arcsecond--scale flux density, with fainter sources being considerably more likely to have parsec--scale emission.  This observation is consistent with the hypothesis that lower--luminosity sources have on average slower radio jets and wider beaming angles, and hence their core emission is less likely to suffer Doppler suppression and more likely to contribute significantly to the overall radio emission when seen from an arbitrary viewing angle.  Finally, \mjive\ has detected 53 variable sources which have considerably increased in flux density since their original FIRST observations, and has shown that previously identified mJy transient sources are more likely to contain a compact component than a typical FIRST source, as expected. 

\acknowledgements  The authors are grateful to J. Morgan for useful suggestions concerning implementation of primary beam corrections, L. Godfrey for useful discussions concerning AGN radio jet models and M. Garrett for useful discussions concerning VLBI detection fractions for FIRST sources.  The National Radio Astronomy Observatory is a facility of the National Science Foundation operated under cooperative agreement by Associated Universities, Inc.  ATD is supported by an NWO Veni Fellowship. 

\bibliographystyle{apj}
\bibliography{deller_thesis}

\end{document}